\documentclass[12pt,english,reprint,aps,prb]{revtex4-1}
\usepackage{mathptmx}
\usepackage{tgheros}
\usepackage[T1]{fontenc}
\usepackage[utf8]{inputenc}
\usepackage{amsmath}
\usepackage{graphicx}
\usepackage{esint}

\makeatletter
\usepackage{xcolor}
\usepackage{soul}
\usepackage[breaklinks,colorlinks,allcolors=blue!80!black]{hyperref}

\renewcommand{\hl}[1]{#1}

\makeatother

\usepackage{babel}
\begin{document}

\title{Pseudocanalization regime for magnetic dark-field hyperlenses}

\author{Taavi Repän}

\email{tarap@fotonik.dtu.dk}

\affiliation{DTU Fotonik, Technical University of Denmark, {\O}rsteds Plads 343, DK-2800 Kongens Lyngby, Denmark}

\author{Andrey Novitsky}

\affiliation{DTU Fotonik, Technical University of Denmark, {\O}rsteds Plads 343, DK-2800 Kongens Lyngby, Denmark}

\author{Morten Willatzen}

\affiliation{DTU Fotonik, Technical University of Denmark, {\O}rsteds Plads 343, DK-2800 Kongens Lyngby, Denmark}

\author{Andrei V Lavrinenko}

\affiliation{DTU Fotonik, Technical University of Denmark, {\O}rsteds Plads 343, DK-2800 Kongens Lyngby, Denmark}
\begin{abstract}
Hyperbolic metamaterials (HMMs) are the cornerstone of the hyperlens, which brings the superresolution effect from the near-field to the far-field zone. For effective application of the hyperlens it should operate in so-called canalization regime, when the phase advancement of the propagating fields is maximally supressed, and thus field broadening is minimized. For conventional hyperlenses it is relatively straightforward to achieve  canalization by tuning the anisotropic permittivity tensor. However, for a dark-field hyperlens designed to image weak scatterers by filtering out background radiation (dark-field regime) this approach is not viable, because design requirements for such filtering and elimination of phase advancement i.e. canalization, are mutually exclusive. Here we propose the use of magnetic ($\mu$-positive and negative) HMMs to achieve phase cancellation at the output equivalent to the performance of a HMM in the canalized regime. The proposed structure offers additional flexibility over simple HMMs in tuning light propagation. We show that in this ``pseudocanalizing'' configuration  quality of an image is comparable to a conventional hyperlens, while the desired filtering of the incident illumination associated with the dark-field hyperlens is preserved.
\end{abstract}


\maketitle

\section{Introduction}

The diffraction limit has been a notorious challenge in a wide range
of applications. \hl{Optical subwavelength imaging} is one particularly
active research direction and throughout the years various solutions
have been proposed to circumvent the diffraction limit. So far practical
results have emerged from a variety of scanning techniques: scanning
near-field optical microscopy (SNOM)~\cite{SNOMreview} and more recently stimulated
emission depletion (STED) microscopy~\cite{STED1,STED2}. Given the intrinsic slowness
of scanning methods, there has always been an interest in alternative
ways to achieve superresolution imaging. With advances in nanofabrication
methods optical metamaterials have become a very promising research
direction. The idea of using \hl{metamaterials for superresolution}
is also not particularly new \textemdash{} Pendry \cite{Pendry_superlens}
has proposed to use a double-negative metamaterial to form a superlens,
which would be able to image details below the diffraction limit.
The superlens itself is somewhat limited in its applicability, as
practical considerations restrict experimental realizations of such
device \cite{Jacob2006}. A few years later an alternative 
approach which avoided double-negative media was proposed \cite{Jacob2006} and
experimentally demonstrated \cite{Liu2007}. This design (the hyperlens)
was instead based on a hyperbolic metamaterial (HMM) structure for superresolution.
The HMM based design allows to avoid the practical challenges of superlenses: unlike double-negative metamaterials fabrication of the HMMs is not so challenging, as they do not rely on resonant
parts. More importantly, due to their non-resonant nature the HMM
structures are less affected by inevitable material losses especially
in the visible \cite{Jacob2006,Yao930,Valentine2008}.

\hl{A hyperbolic medium} is an anisotropic medium, where the ordinary
and extraordinary permitivitties have opposing signs. Effectively,
the medium is metallic in one direction and dielectric in others.
In solving this system for plane wave propagation the resulting dispersion
relation shows that waves with arbitrarily high wavevectors are allowed
to propagate inside the medium {[}i.e. the isofrequency surface $\omega=f\left(k_{x},k_{y},k_{z}\right)=\mathrm{const}$
is unbounded{]} \cite{Smith2003,Liu2008,Poddubny2013}. This is in
contrast to conventional media, where only a limited range of waves
can propagate, while the rest are evanescent. The evanescent waves
are highly localized and are only accessible by near-field probes
(such as SNOM) \cite{NovtonyNanooptics}. This filtering of high-k
waves in the far field results in the diffraction limit. The ability
of hyperbolic media to carry these waves allows for superresolution
imaging by subverting the diffraction limit.

To facilitate a straightforward imaging process, the hyperlens should
be designed such that the fields propagate through the device with
minimal distortion \cite{Lu2012,Xiong2009,Ikonen2007,Rho2010}. This
is achieved when the HMM is engineered to have
the permittivity tensor {[}$\hat{\varepsilon}=\mathrm{diag}(\varepsilon_x,\varepsilon_y,\varepsilon_x)${]} feature either the epsilon-near-zero
($\varepsilon_x \approx 0$) or epsilon-near-pole ($|\varepsilon_y| \gg 1$) components
(i.e. the HMM is operated in the so-called canalization regime). As a consequence the
plane wave field components acquire very little phase, meaning that
the image is effectively \textquotedblleft canalized\textquotedblright{}
through the medium, exhibiting very little broadening or diffraction.

Natural objects for superresolution imaging would be biological samples in scale of a few hundred nanometers,
which are relatively weakly scattering (compared to plasmonic particles, for example). As the usual design of the 
hyperlens carries both incident and scattered waves the available contrast 
is not enough in case of weakly scattering objects. To facilitate
imaging of weakly scattering subwavelength objects, \hl{dark-field hyperlens}
designs were proposed \cite{Benisty2012,Repan2015}. For example, by appropriately choosing signs
of the permittivity tensor components~\cite{Repan2015} a hyperbolic medium can be engineered
to filter out waves with long effective wavelength (small wavenumber).
This mode of operation (termed type-II HMM) can be used as a basis for
a dark field hyperlens~\cite{Repan2015}.
However, the design based on type-II HMM suffers from the lack of
canalization regime and reduced device performance. 

In this paper, we propose a new method for circumventing the diffraction
limit using a canalization regime. In doing this, we provide a detailed
discussion of wave propagation in hyperbolic materials with special
emphasis to canalization solutions. We show from the propagation equations
that the image broadening in hyperbolic media has two different contributions:
absorption (determined by material losses) and phase accumulation
(determined by the dispersion relation). The absorption term is difficult
to circumvent, but the phase propagation may be decreased (for example
by employing the canalization regime). However, we show that the canalization
regime for a homogeneous hyperbolic medium is fundamentally incompatible
with dark-field imaging (based on low-k filtering). Relaxing the requirement
of a homogeneous medium leads to the idea of a ``pseudocanalization''
regime, where instead of a single medium we aim to use two complementary
media to compensate phase advances of each other (allowing for reduced
broadening), while keeping the low-k filtering properties necessary
for dark-field imaging. This complementary medium can be realized
using a $\mu$-negative HMM, extending the idea for isotropic media
from Ref.~\onlinecite{Alu2003}.

We start by outlining basic theory of light propagation in HMMs in
Section~\ref{sec:Basic-theory}. We follow with discussion about
the canalization regime in hyperbolic media (Section~\ref{sec:Canalization-regime}).
In Section~\ref{sec:Phase-compensation} we propose and discuss the
idea of pseudocanalization by using $\mu$-negative HMMs for phase
compensation. We demonstrate applicability of the idea in Section
\ref{sec:Improved-dark-field-hyperlens}, with particular focus on
improving a dark-field hyperlens.

\section{Basic theory\label{sec:Basic-theory}}

\subsection{Propagation of waves and dispersion equation}

To study propagation of plane waves in homogeneous medium we consider
an angular spectrum representation~\cite{NovtonyNanooptics} of the
fields, where the initial electric fields at $y=0$ are decomposed
with the Fourier transform
\begin{equation}
\mathbf{E}\left(k_{x},y=0\right)=\frac{1}{2\pi}\intop\mathbf{E}\left(x,y=0\right)\exp\left(-ik_{x}x\right)dx\,.
\end{equation}

Propagated fields after a distance $y$ can then be calculated with

\begin{equation}
\mathbf{E}\left(k_{x},y\right)=\mathbf{E}\left(k_{x},y=0\right)\exp\left(ik_{y}y\right)\,,\label{eq:kx_propagation}
\end{equation}
where $k_{y}$ is the propagation constant (wavevector component along
the $y$-axis). Using an inverse Fourier transformation, we find:

\begin{equation}
\mathbf{E}\left(x,y\right)=\intop\mathbf{E}\left(k_{x},y\right)\exp\left(ik_{x}x\right)dk_{x}\,.\label{eq:propagation}
\end{equation}

To apply Eq.~(\ref{eq:kx_propagation}) for HMMs we need the expression
for the propagation constant $k_{y}$ in an anisotropic medium. We
assume an anisotropic permittivity {[}$\hat{\varepsilon}=\mathrm{diag}\left(\varepsilon_{x},\varepsilon_{y},\varepsilon_{x}\right)${]}
and an isotropic permeability ($\mu$). Assuming next a plane wave
solution for Maxwell's equations the dispersion relation for extraordinary waves becomes
(see details in Appendix \ref{sec:Derivation-of-dispersion})

\begin{equation}
\frac{k_{x}^{2} + k_z^2}{\varepsilon_{y}}+\frac{k_{y}^{2}}{\varepsilon_{x}}=\mu k_{0}^{2}\,,\label{eq:dispersion}
\end{equation}
which describes propagation of plane waves through the medium. In this paper we assume propagation in the $x$-$y$ plane, i.e. $k_z=0$. We
can solve Eq.~(\ref{eq:dispersion}) to yield the propagation constant
in the $y$-direction

\begin{equation}
k_{y}=\pm\sqrt{\varepsilon_{y}\varepsilon_{x}\left(\varepsilon_{y}k_{0}^{2}\mu-k_{x}^{2}\right)}/\varepsilon_{y}\,.\label{eq:propagation_constant}
\end{equation}

The sign of $k_{y}$ can be established using the Poynting vector
direction -- we are interested in waves propagating towards the positive
$y$ direction. The $y$-component of the Poynting vector in our case
can be written as

\begin{equation}
S_{y}=\frac{\left|H_{z}\right|^{2}}{2\omega}\mathrm{Re}\left(\frac{k_{y}}{\varepsilon_{0}\varepsilon_{x}}\right)\,.
\end{equation}
To have propagation in the positive $y$ direction, the sign of $k_{y}$
must be chosen to have $S_{y}>0$.

In the general case the propagation constant has both real and imaginary
parts $k_{y}=k_{y}'+\mathrm{i}k_{y}''$, where the real (imaginary)
part describes phase accumulation (attenuation).

\subsection{Propagation in hyperbolic media}

\begin{figure}
\begin{centering}
\includegraphics{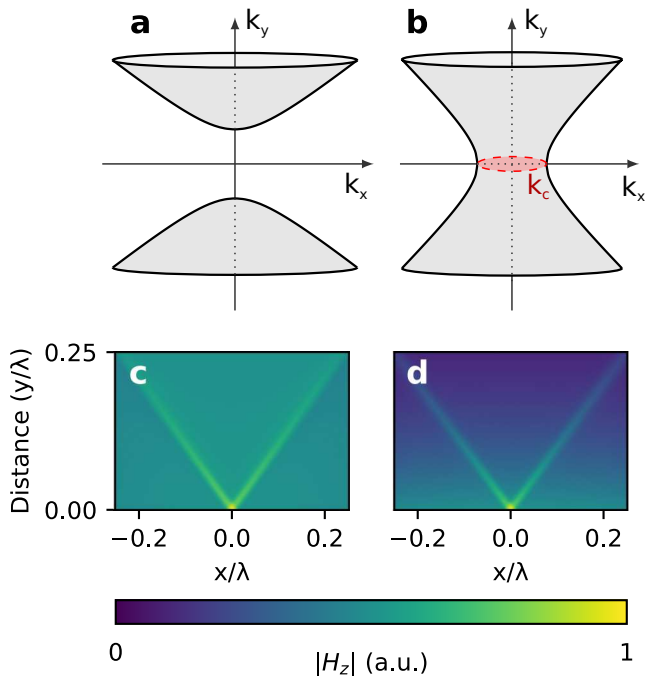}
\par\end{centering}

\caption{\label{fig:hmm-types}(a,b) Isofrequency surfaces for type-I (a) and
type-II (b) hyperbolic dispersions. The surfaces are unbounded, allowing
propagation of waves with arbitrarily large $k_{x}$. However, in
type-II HMM (b) the low-k waves ($k_{x}<k_{c}$) are not propagating
waves (marked with red circle). (c,d) Propagation of fields from a
point source along with uniform background fields in type-I HMM $\varepsilon_x=1+0.05\mathrm{i}$, $\varepsilon_y=-1+0.05\mathrm{i}$ (c)
and type-II HMM $\varepsilon_x=-1+0.05\mathrm{i}$, $\varepsilon_y=1+0.05\mathrm{i}$ (d). Important feature of type-II HMM is that the
background fields are attenuated along propagation.}
\end{figure}

In hyperbolic media components $\varepsilon_{x}$, $\varepsilon_{y}$
have different signs. In this case the isofrequency contour of Eq.~(\ref{eq:dispersion})
will yield a hyperboloid, as shown in Fig.~\ref{fig:hmm-types}(a,b).
Two different configurations can be distinguished: we designate the
case of $\mu\varepsilon_{y}<0<\mu\varepsilon_{x}$ as type-I hyperbolic
dispersion and the case of $\mu\varepsilon_{x}<0<\mu\varepsilon_{y}$
an type-II hyperbolic dispersion~\cite{Shekhar2014}. We note that
asymptotic behavior for large $k_{x}$ is the same for type-I and
type-II hyperbolic media: 
\begin{equation}
k_{y}\propto\pm k_{x}\sqrt{\left|\varepsilon_{x}\right|/\left|\varepsilon_{y}\right|}\,.
\end{equation}
This indicates that $k_{y}$ will stay real for arbitrary high $k_{x}$
and consequently the high-k waves are always propagating waves in
the hyperbolic medium. This is different from conventional media,
where the propagation constant for high-k waves turns fully imaginary,
signifying evanescent nature of the fields \cite{NovtonyNanooptics}.

The differences between type-I and type-II hyperbolic media become
apparent for low-k waves. The propagation constant at $k_{x}=0$ becomes 

\begin{equation}
k_{y}\left(0\right)=\pm\sqrt{\varepsilon_{x}k_{0}^{2}\mu}\,.\label{eq:ky(0)}
\end{equation}

For type-II HMM $\varepsilon_{x}\mu<0$ so the propagation constant
becomes imaginary, i.e. these waves are evanescent (for conventional
medium) or amplified (in case of gain medium). The transition point
$k_{c}$ between low-k and high-k waves can be seen from Eq.~(\ref{eq:propagation_constant})
:

\begin{equation}
k_{c}^{2}=\varepsilon_{y}\mu k_{0}^{2}\,,\label{eq:k_cutoff}
\end{equation}
meaning that in a type-II HMM only waves with $\left|k_{x}\right|\geq k_{c}$
(high-k) will be propagating, whereas waves with $\left|k_{x}\right|<k_{c}$
(low-k) will be evanescent. As the subwavelength details of an image
are mostly contained in high-k waves,while the background field is
transported by low-k waves, this filtering can be used to design a
dark-field version of the hyperlens~\cite{Repan2015}. We use Eq.~(\ref{eq:kx_propagation}) to calculate the propagation of the fields through the hyperbolic medium to illustrate the key difference between type-I and type-II HMMs, namely the filtering of background radiation in type-II hyperbolic medium. Figure~\ref{fig:hmm-types}(c,d)
shows that the type-II HMM filters out background radiation, unlike
the type-I HMM.

\section{Canalization regime\label{sec:Canalization-regime}}

One of the early proposals for a hyperlens was based on a metamaterial
consisting of a wire medium \cite{Belov2006_wires}. In such a medium
the modes propagating in individual wires transport pixels of the
image. In other words, the image is ``canalized'' from the inner
to outer interface, giving rise to the name of this mode of operation.
This is important for imaging purposes, as the fields propagate with
minimal distortion. However, such operation is not limited to wire
media: similar operation can be obtained with various configurations
of hyperbolic metamaterials \cite{Salandrino2006,Jacob2007,Lee2007}.
In general, a hyperbolic medium approaches the canalization regime
as either $\varepsilon_{x}$ approaches zero and/or $\varepsilon_{y}$
approaches infinity. In these limits the propagation constant becomes
independent of $k_{x}$ and from Eq.~(\ref{eq:propagation}) it follows
that fields will propagate in an undistorted manner:
\begin{equation}
\mathbf{E}(x,y)=\exp\left(ik_{y}y\right)\mathbf{E}\left(x,0\right)\,.
\end{equation}
As a result, fields are ``canalized'' through the medium. For superresolution
imaging this regime is strongly desirable, as it is vital for a distortion-free
image. Most hyperlens designs proposed so far utilize the canalization
regime. For a detailed discussion on HMMs where $\varepsilon_x\approx0$ see Ref.~\onlinecite{Castaldi2012}.

The canalization regime also implies minimal broadening of the image.
For waves propagating in $y$-direction, we can estimate broadening
using the Poynting vector components: 

\begin{equation}
\frac{S_{x}}{S_{y}}=\frac{\mathrm{Re}\left(k_{x}/\varepsilon_{y}\right)}{\mathrm{Re}\left(k_{y}/\varepsilon_{x}\right)}\,,\label{eq:Poynting_spread}
\end{equation}
which in canalization limit approaches zero (for the lossless case).
This shows that in a canalizing system the fields propagate directly
in $y$-direction, i.e. there is no broadening. In lossy systems the
broadening will have two contributions: one arises from attenuation
of waves which affects high-k waves more and thus narrows the spectrum
in the reciprocal space. This corresponds to broadening in the real
space. Furthermore, spread of the Poynting vector {[}as per Eq.~(\ref{eq:Poynting_spread}){]}
also causes broadening. This spread is linked to the phase accumulation
of propagating fields and can therefore, in principle, be compensated.
For comparison, compensating for attenuation losses is impossible
without using gain media.

It is important to note that the distinction between a type-I and
type-II HMM disappears in the canalization regime. Firstly, considering
the limit where $\varepsilon_{y}\to\infty$, we see from Eq.~(\ref{eq:propagation_constant})
that $k_{y}\approx\sqrt{\varepsilon_{x}\mu}k_{0}$ (for large $\varepsilon_{y}$),
which means that all fields will be either propagating or evanescent,
depending on the sign of $\varepsilon_{x}\mu$. As a consequence,
there is no distinction between low-k and high-k waves. In other case
($\varepsilon_{x}\to0$) we see that the distinction between low-k
and high-k waves is unaltered {[}the cut-off point $k_{c}$ is independent
of $\varepsilon_{x}$, as per Eq.~(\ref{eq:k_cutoff}){]}. However,
from Eq.~(\ref{eq:propagation_constant}) it follows that the propagation
constant $k_{y}$ scales with $\sqrt{\varepsilon_{x}}$. This means
that as the system moves closer to the canalization regime the attenuation
constant ($\mathrm{Im}\,k_{y}$) is reduced therefore nullifying the
low-k filtering effect.

\begin{figure}
\begin{centering}
\includegraphics{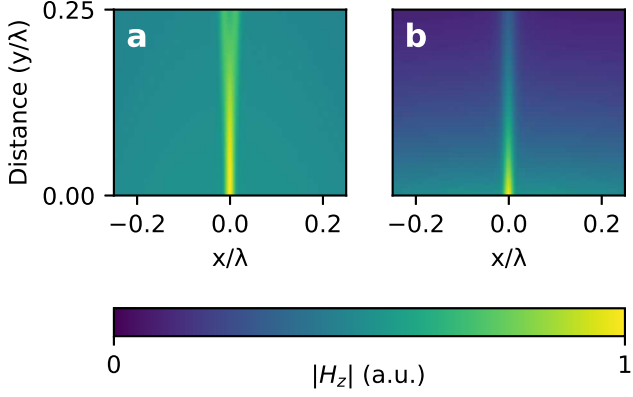}
\par\end{centering}

\caption{\label{fig:pseudocanalized-filtering}Propagation of fields through
a canalizing (a) and a pseudocanalizing (b) type-II HMM slab. 
The canalizing HMM parameters are $\varepsilon_x=\left(-1+0.05\mathrm{i}\right)/20^2$, $\varepsilon_y=1+0.05\mathrm{i}$. To achieve pseudocanalization a slab of type-II HMM [same as in Fig.~\ref{fig:hmm-types}(d)] is combined with a compensating slab [given by Eqs.~(\ref{eq:pc_condition_1}) and~(\ref{eq:pc_condition_2}), with $\mu^{(2)}=-1$] of equal thickness.
Initial field is the superposition of the field of a narrow point-like source
and a uniform background field. Note that unlike the noncanalizing
type-II HMM {[}Fig.~\ref{fig:hmm-types}(d){]}, there is no attenuation
of the background fields in (a). However, in the pseudocanalizing
system (b) the background filtering properties are restored. 
}
\end{figure}

The lack of low-k filtering is shown in Fig.~\ref{fig:pseudocanalized-filtering}(a),
which shows that in the canalization regime the type-II HMM does not
attenuate background fields {[}compare against Fig.~\ref{fig:hmm-types}(d){]}.
As the background fields are not attenuated, the canalization regime
is not applicable for dark-field imaging. However, working beyond
the canalization regime degrades the image quality of hyperlens, creating
additional challenges \cite{Repan2015}. Therefore, it would be beneficial
to achieve canalization-like behavior while still maintaining the
low-k filtering properties.

\begin{figure}
\begin{centering}
\includegraphics{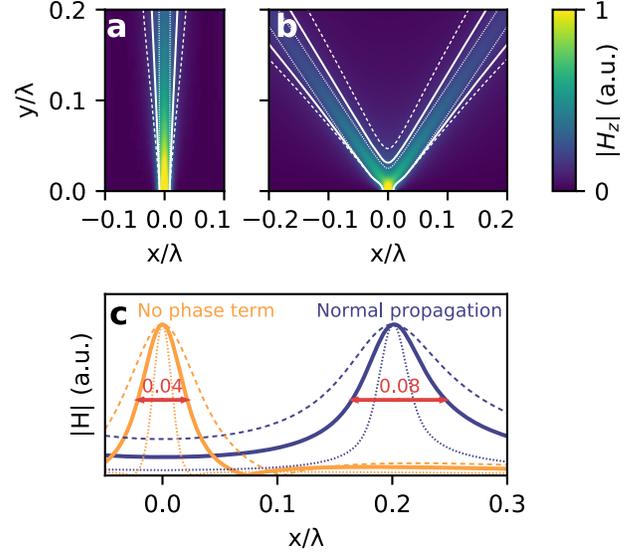}
\par\end{centering}

\caption{\label{fig:broadening}Broadening of the fields during propagation
through type-II HMM ($\varepsilon_{x}=-1+\mathrm{i}\gamma$, $\varepsilon_{y}=1+\mathrm{i}\gamma$,
$\gamma=0.1$). Propagation of fields through the slab, for normal
type-II slab (b) and for phase-less propagation, i.e. $\mathrm{Re}\,k_{y}=0$
(a). FWHM of the beam with various losses is indicated by the solid
($\gamma=0.1$), dashed ($\gamma=0.2$) and dotted ($\gamma=0.05$)
lines. (c) Fields for both cases (normal and phase-less propagation)
after propagation distance $z=0.2$. The dashed and dotted lines indicate
the same loss factors $\gamma$. The results indicates that the phase
term is responsible for having $\sim2\times$ increase in FWHM.}
\end{figure}

From the propagation equation {[}Eq.~(\ref{eq:kx_propagation}){]}
we note that the effect of the propagation constant $k_{y}$ can be
split into the real ($k_{y}'$) and imaginary ($k_{y}''$) parts.
The real part will yield a phase term {[}$\exp\left(ik_{y}'y\right)${]},
while the other term yields an attenuating term {[}$\exp\left(-k_{y}''y\right)${]}.
The latter affects the high-k waves more, causing narrowing of the
wavevector spectrum (broadening in the image space). This broadening
is shown in Fig.~\ref{fig:broadening}(a), where only the attenuation
term is taken into account. However, the phase term will also contribute
to broadening, as seen in Fig.~\ref{fig:broadening}(b), where both
phase and attenuation terms are considered. Comparing the two cases
{[}Fig.~\ref{fig:broadening}(c){]}, we see that phase term causes
additional broadening. As the canalization regime implies minimal
phase distortions, the additional broadening term is suppressed in
this regime.

\section{Phase compensation with $\mu$-negative HMM\label{sec:Phase-compensation}}

The key property of the canalization regime (in the ideal limit) is
that fields propagate with constant phase accumulation {[}i.e. $\mathrm{Re}\left(k_{y}\right)=\mathrm{const}${]}.
However, the canalization regime is not a strict prerequisite for
having no phase accumulation. The identical result could be achieved
by replacing the homogeneous HMM medium with two different HMM slabs with complementary dispersion, such that 
\begin{equation}
d_1 k_{y}^{(1)}\left(k_{x}\right)=-d_2 k_{y}^{(2)}\left(k_{x}\right)\,,\label{eq:pseudocanalization_condition}
\end{equation}
where $d_1$, $d_2$ are the thicknesses of two slabs. In most of the calculations here we assumed $d_1=d_2$, but in general thicknesses can be varied to allow more freedom in engineering suitable permittivity and permeability properties.
Assuming full impedance matching (ie. no reflections) between the
slabs, the propagated fields will have no distortions in the phase
or amplitude. In this case the two slabs form effectively a canalizing
system.

To proceed with calculations, we assume two lossless hyperbolic media:
the first medium has $\mu^{(1)}=1$, whereas for the second medium we require
$\mu^{(2)}<0$. We can repeat Eq.~(\ref{eq:pseudocanalization_condition})
for the two defining cases, firstly for low-k waves ($k_{x}=0$)

\begin{equation}
d_1 \mathrm{sgn}\left(\varepsilon_{y}^{(1)}\right)\sqrt{\varepsilon_{x}^{(1)}}
= -d_2 \mathrm{sgn}\left(\varepsilon_{y}^{(2)}\right)\sqrt{\mu^{(2)} \varepsilon_{x}^{(2)}}
\,,\label{eq:PC_eq_1}
\end{equation}
and then for high-k waves by taking the limit where $k_{x}$ approaches
infinity:

\begin{equation}
d_1 \sqrt{-\varepsilon_{x}^{(1)}\varepsilon_{y}^{(1)}}k_{x}/\varepsilon_{y}^{(1)}=-d_2 \sqrt{-\varepsilon_{x}^{(2)}\varepsilon_{y}^{(2)}}k_{x}/\varepsilon_{y}^{(2)}\,.\label{eq:PC_eq_2}
\end{equation}

Solving Eqs. (\ref{eq:PC_eq_1}) and (\ref{eq:PC_eq_2}) yields conditions
for two slabs:

\begin{eqnarray}
\mu^{(2)} \varepsilon_{x}^{(2)} & = & \varepsilon_{x}^{(1)} d_1^2/d_2^2\label{eq:pc_cond_lossless_1}\\
\mu^{(2)} \varepsilon_{y}^{(2)} & = & \varepsilon_{y}^{(1)}\,.\label{eq:pc_cond_lossless_2}
\end{eqnarray}

Although we derived the relations based only on two cases ($k_{x}=0$
and $k_{x}=\infty$), it is easy to verify that the phase-matching
condition {[}Eq.~(\ref{eq:pseudocanalization_condition}){]} holds
for all $k_{x}$. In case of lossy media, we limit the discussion
to media where $\mathrm{Im}\left(\varepsilon\right)>0$, i.e. we neglect
gain media. In lossy media we only require the real part of Eq.~(\ref{eq:pseudocanalization_condition})
to hold. However, we do assume $d_1 \mathrm{Im}\,k_{y}^{(1)}=d_2 \mathrm{Im}\,k_{y}^{(2)}$,
so that we reach the following conditions for complex permitivitties:
\begin{align}
\mu^{(2)} \varepsilon_x^{(2)} &= \varepsilon_x^{(1)*} d_1^2/d_2^2 \label{eq:pc_condition_1}\\
\mu^{(2)} \varepsilon_y^{(2)} &= \varepsilon_y^{(1)*} \label{eq:pc_condition_2}
\end{align}

The conditions above ensure that the phase accumulation is canceled
even in lossy media. However, due to losses the fields will not stay
unmodified --- the plane wave components of the image will be attenuated,
where attenuation factor $\mathrm{Im}\left(k_{y}\right)$ depends
on $k_{x}$. As different plane wave components experience different
attenuation, this will result in broadening of the image in real space.
However, as discussed in the previous section, the broadening in HMMs
is caused both by phase and attenuation terms. In a pseudocanalizing
system the contribution from the phase term is eliminated. Figure~\ref{fig:broadening}(c)
shows that even when considering the losses, the broadening is greatly
reduced in a pseudocanalizing system.

It is important to stress that the constituent media are not required
to be in the canalization regime. This means that we can have two
complementary type-II HMM slabs (both exhibiting low-k filtering)
and combine them in the pseudocanalizing system with dark-field operation.

From the impedance for oblique incidence ($\gamma=k_{y}/\varepsilon_{x}k_{0}$),
conditions for slab permitivitties {[}Eqs. (\ref{eq:pc_cond_lossless_1})
and (\ref{eq:pc_cond_lossless_2}){]} and wavenumbers Eq.~(\ref{eq:pseudocanalization_condition})
we see that the pseudocanalizing slabs are impedance matched, for
the lossless case. As the designs of practical interest are limited
to low loss regime, we continue to neglect reflections for lossy system
as well. We later show with full-wave simulations that reflections
do not significantly alter performance of the device. Therefore
propagation of the initial fields through the two slabs can be written in accordance with Eq.~(\ref{eq:kx_propagation}) as

\begin{equation}
\mathbf{E}\left(k_{x},y\right)=\begin{cases}
\mathbf{E}\left(k_{x},0\right)\exp\left(ik_{y}^{(1)}y\right) & y<h\\
\mathbf{E}\left(k_{x},0\right)\exp\left(ik_{y}^{(1)}h\right)\exp\left(ik_{y}^{(2)}\left(y-h\right)\right) & y>h
\end{cases}\,,\label{eq:calculations}
\end{equation}
where $h$ is the thickness of the first slab.
In Fig.~\ref{fig:pseudocanalized-filtering}(b) we use this to show operation of such pseudocanalizing system. We see that the original image is transmitted with minimal distortion (similar to the canalizing medium), except for the background, which is strongly attenuated. 

\begin{figure}
\begin{centering}
\includegraphics{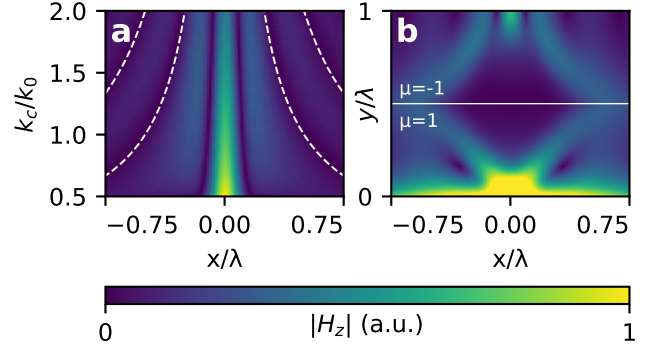}
\par\end{centering}

\caption{(a) Magnetic fields after propagation of a fixed distance ($y=\lambda$,
$h=\lambda/2$) of pseudocanalizing type-II HMM, with varying low-k
cutoff $k_{c}$. Dashed lines indicate expected zeros at $n\pi/k_{c}$,
present due low-k filtering. Material parameters are $\varepsilon_{x}=-k_{c}^{2}/k_{0}^{2}$,
$\varepsilon_{y}=k_{c}^{2}/k_{0}^{2}$. Initial fields are $H_{z}=\exp\left(-x^{2}/\Delta^{2}\right)$,
where $\Delta=\lambda/10$. (b) Operation of pseudocanalizing system
for $k_{c}=1.5k_{0}$. The figures are calculated using Eq.~(\ref{eq:calculations}).
\label{fig:filtering-distrortion}}
\end{figure}

\label{Discussion:distortions-in-type-2}However, as shown in Fig.~\ref{fig:filtering-distrortion},
the image from the type-II pseudocanalizing system is not completely
distortion-free: attenuation of low-k waves behaves as a high-pass
filter for the image, somewhat reducing the image quality. The $\exp\left(ik_{y}y\right)$
term in the propagation equation {[}Eq.~(\ref{eq:propagation}){]}
can be approximated as a high-pass filter with the cut-off at $k_{c}$.
Assuming a point source, we can use the Fourier transform of a rectangular
function to approximate the filtered image as

\begin{equation}
\mathbf{E}\left(x,y\right)\approx\mathbf{E}_{0}\frac{2}{x}\sin\left(k_{c}x\right)\,.
\end{equation}
We see that due to filtering the image will have additional zeros
at $n \pi/k_{c}$ (with $n=0,\pm1,...$). This is made worse with
increasing $k_{c}$ , as the image will develop more sidelobes.

Finally, it's worth pointing out that the phase compensation could
be achieved in limited cases without $\mu$-negative materials as
well. It is easy to show that phase compensation condition for high-k
waves {[}Eq.~(\ref{eq:PC_eq_2}){]} will have the same form even
when $\mu^{(2)}=\mu^{(1)}=1$. This means that phase compensation
can be achieved with a medium where

\begin{eqnarray}
\mathrm{sgn}\left(\varepsilon_{y}^{(1)}\right) & = & -\mathrm{sgn}\left(\varepsilon_{y}^{(2)}\right)\,,\\
\varepsilon_{x}^{(1)}/\varepsilon_{y}^{(1)} & = & \left( \varepsilon_{x}^{(2)} d_1^2\right) / \left( \varepsilon_{y}^{(2)} d_2^2 \right)\,.
\end{eqnarray}
This only partially solves the problem since for low-k waves the phase
is not canceled {[}Eq.~(\ref{eq:PC_eq_1}) is not satisfied{]}. For
example, in superresolution imaging applications most of the energy
will be carried by waves with $k_{x}$ near $k_{c}$, both due to
their evanescent nature in the medium outside the hyperlens and also
due to material losses having stronger effect on waves with higher
$k_{x}$. Since the phase of waves near $k_{c}$ is not compensated
the image will experience significant distortion.

\section{Pseudocanalization in a cylindrical hyperlens\label{sec:Improved-dark-field-hyperlens}}

We now extend the discussion from planar to cylindrical geometry as
is the case with hyperlens structures. For cylindrical geometry the
hyperbolic permittivity is given as $\hat{\varepsilon}=\mathrm{diag}\left(\varepsilon_{\theta},\varepsilon_{r},\varepsilon_{\theta}\right)$.
A simple plane wave analysis suffices for a qualitative understanding.
The detailed numerics is carried out using a full-wave EM analysis.
From geometric principles it follows straightforwardly that the image
is expanded by a factor of
\begin{equation}
M=\frac{r_{2}}{r_{1}}\,,
\end{equation}
where $r_{1}$ is the initial radius from which the fields start
propagating (inner surface of the hyperlens) and $r_{2}$ is the final
radius (outer surface of the hyperlens). In angular spectrum representation
this corresponds to a compression of $k_{\theta}$ as the wave propagates from the initial value $k_\theta'$ to

\begin{equation}
k_{\theta}=k_{\theta}'/M\,,
\end{equation}
where $k_{\theta}'$ is the initial value. To counteract effects of
magnification, the second medium (compensation medium) should be scaled.
As we show in Appendix~\ref{sec:Cylindrical-theory...}, $\varepsilon_{r}^{(2)}$
should be scaled by the total magnification of the hyperlens:

\begin{equation}
\varepsilon_{r}^{(2)}=\varepsilon_{r}^{\prime(2)}/M\,.\label{eq:er_scaling}
\end{equation}

To demonstrate the concept, we have chosen relatively simple material parameters for the HMMs, of the form $\varepsilon_{x,y}=\pm 1 + \gamma \mathrm{i}$. This allows us to capture the effects related to losses, while keeping the number of free parameters minimal. However, we point out the choice is only done to demonstrate a simple analysis, as the conditions in Eqs.~(\ref{eq:pc_condition_1}) and~(\ref{eq:pc_condition_2}) are general and are not limited to the simplifed paramaters used here.

\begin{figure}
\begin{centering}
\includegraphics{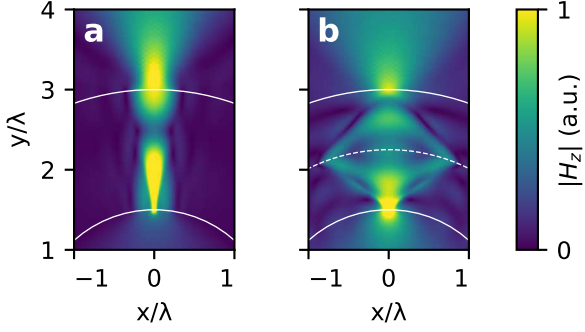}
\par\end{centering}

\caption{Simulations of a radiation of a point source close to the inner interface
imaged by a bright field hyperlens {[}$\varepsilon_{\theta}=0.2+0.02\mathrm{i}$,
$\varepsilon_{r}=-5+2.20\mathrm{i}${]}, operating in the
canalization regime (a). Similar bright-field hyperlens, but using
pseudocanalizing structure consisting of two HMM layers {[}$\varepsilon_{\theta}^{(1)}=1+0.05\mathrm{i}$,
$\varepsilon_{r}^{(1)}=-1+0.05\mathrm{i}$, $\varepsilon_{\theta}^{(2)}=\varepsilon_{\theta}^{(1)*}$,
$\varepsilon_{r}^{(2)}=\varepsilon_{r}^{(1)*}/2${]} (b). For all
structures the inner radius is $r_{1}=1.5\lambda$ and the outer radius
$r_{2}=3\lambda$. Dashed line in (b) shows split between $\mu$-positive
and $\mu$-negative media. \label{fig:bright-hyperlens}}
\end{figure}

We start by demonstrating the concept with a bright field hyperlens
{[}Fig.~\ref{fig:bright-hyperlens}(a){]}, based on a type-I HMM.
This allows us to directly compare canalizing and pseudocanalizing
operations in cylindrical geometry. We show that the pseudocanalizing
system works in cylindrical geometry too {[}Fig.~\ref{fig:bright-hyperlens}(b){]},
demonstrating applicability for hyperlens devices. Figure \ref{fig:bright-hyperlens}
is obtained by full-wave simulations (using\textsc{ COMSOL Multiphysics}
v5.1), i.e. reflections from interfaces are taken into account. Standing
waves originated from these reflections are seen on the figures as
the modulated intensity along the direction of propagation.

\begin{figure}
\begin{centering}
\includegraphics{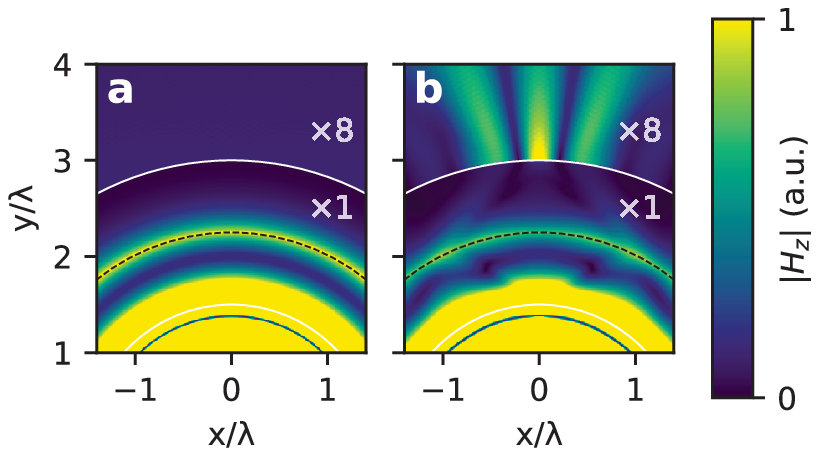}
\par\end{centering}

\caption{\label{fig:dark-hyperlens}Simulations demonstrating functioning of
a dark-field hyperlens with two complementary HMM layers operating
in pseudocanalized regime {[}$\varepsilon_{\theta}^{(1)}=-1+0.05\mathrm{i}$,
$\varepsilon_{r}^{(1)}=1+0.05\mathrm{i}$, $\varepsilon_{\theta}^{(2)}=1+0.05\mathrm{i}$,
$\varepsilon_{r}^{(2)}=\left(-1+0.05\mathrm{i}\right)/2$, $\mu^{(1)}=1$, $\mu^{(2)}=-1${]}.
Simulations without (a) and with (b) a small subwavelength scatterer
show that the incoming background radiation is filtered, whereas scattered
fields from the weakly scattering dielectric object pass through the
system.}
\end{figure}

Similar considerations hold for a dark-field hyperlens (based on a
type-II HMM): the scaling for the compensation layer follows the same
relation {[}Eq.~(\ref{eq:er_scaling}){]}. Simulation results for
the dark-field hyperlens are shown in Fig.~\ref{fig:dark-hyperlens}:
the background radiation is still filtered out (dark-field operation
is preserved), while scattered fields from the small dielectric
object are passing through the device. As discussed in Section~\ref{Discussion:distortions-in-type-2},
the filtering of low-k waves affects the image so that dark-field
operation comes at the cost of image quality. By reducing the low-k
cutoff ($k_{c}$) the image quality is improved, as we discussed in
case of the flat geometry {[}see Fig.~\ref{fig:filtering-distrortion}(a){]}.

\begin{figure}
\begin{centering}
\includegraphics{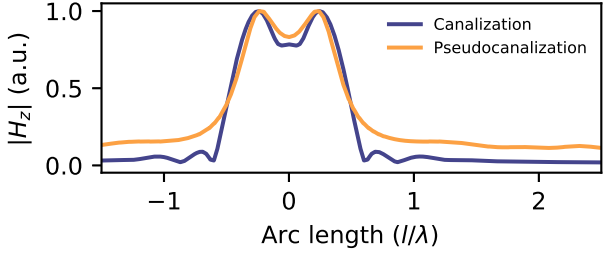}
\par\end{centering}

\caption{\label{fig:resolution-limit-bright}Magnetic field norm along the outer interface of the hyperlens due to two point sources. The two sources near the inner interface are separated by $\lambda/4$. The blue line shows behavior of canalizing bright-field hyperlens [Fig.~\ref{fig:bright-hyperlens}(a)] while the orange line shows the corresponding case for a pseudocanalizing hyperlens [Fig.~\ref{fig:bright-hyperlens}(b)].}
\end{figure}

The hyperlens resolution is determined by two factors: magnification and broadening. The fields on the outer interface should be imaged by far-field optics and hence must be above the diffraction limit. Given hyperlens with magnification $M$ the resolution limit is then $\lambda/2M$. In our case, the hyperlens geometry used has $2\times$ magnification leading to resolution of $\lambda/4$. However, the broadening inside the HMM is also an important consideration: if the beams originating from a point source are broadened too much, they will overlap in the output. In such a case the resolution will be limited by the broadening of the beams. In Fig.~\ref{fig:resolution-limit-bright} we show that both canalizing and pseudocanalizing bright-field hyperlenses have comparable performance in the limiting case of $\lambda/4$ separation. In Fig.~\ref{fig:resolution-limit-bright} we show that both canalizing [Fig.~\ref{fig:bright-hyperlens}(a)] and pseudocanalizing bright-field hyperlens [Fig.~\ref{fig:bright-hyperlens}(b)] have small enough broadening so that the required resolution $\lambda/4$ can be reached. However, for the hyperlens used here, if a better resolution is needed, then losses must be reduced, otherwise the broadening will be a limiting factor.

\begin{figure}
\begin{centering}
\includegraphics{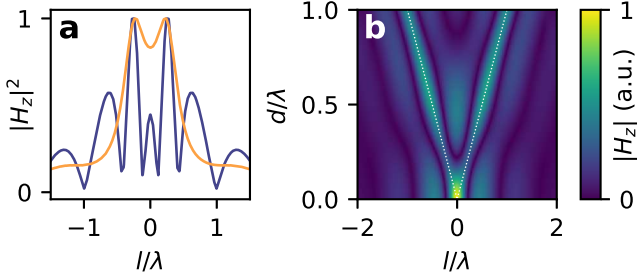}
\par\end{centering}

\caption{\label{fig:resolution-limit-dark}(a) Magnetic field norm along the outer interface of a pseudocanalizing bright-field hyperlens (orange line) and a pseudocanalizing dark-field hyperlens (blue ling). The two point sources are separated by $\lambda/4$. (b) Magnetic field norm along the outer interface as a function of the point source separation $d$. The dotted lines show location of the main peaks from the two point sources.}
\end{figure}

In Fig.~\ref{fig:resolution-limit-dark}(a) we compare the performance of the pseudocanalizing bright-field [Fig.~\ref{fig:bright-hyperlens}(b)] and the dark-field hyperlens (Fig.~\ref{fig:dark-hyperlens}). At the resolution limit the pseudocanalizing dark-field hyperlens offers competitive performance with respect to the bright-field hyperlens. We even see an edge-enhancing behavior associated with high-pass filtering due to filtering low-k waves, which enhances the contrast between the two peaks. 

However, as shown in Fig.~\ref{fig:resolution-limit-dark}(b) the behavior of a dark-field hyperlens is not straightforward. Although (magnification-limited) resolution of $\lambda/4$ is reached with relative ease, the sidelobes associated with low-k filtering are present and could be problematic for some configurations. In our case, the worst case happens for a source separation of $0.36\lambda$ where the sidelobes of two sources interfere constructively. Nonetheless, here the effect is not strong enough to pose serious problems: the ratio between the main peak and the highest sidelobe is about 1.2. This is comparable to the worst case performance of the bright-field pseudocanalizing hyperlens, shown in Fig.~\ref{fig:resolution-limit-dark}(a), where the ratio between the peak and the valley between the peaks is also around 1.2.

\section{Conclusions}

We have shown that the effect of the canalization regime can be understood
as propagation of fields without phase accumulation. This suppression
of phase term minimizes the signal broadening in the HMM and prevents distortions
of the image. We have extended the idea of the canalization regime from homogeneous
hyperbolic media to systems consisting of two complementary hyperbolic
slabs (pseudocanalizing system), where the phase propagation in the
slabs cancels each other so that the propagated fields have no any additional
phase term. Unlike a canalizing system, a pseudocanalyzing system
allows to sustain the dark-field imaging properties along with the
image quality comparable to a canalizing HMM.

This idea of pseudocanalization also applies for cylindrical geometries,
i.e. for typical hyperlens designs. Magnification stemming from cylindrical
geometry creates some new considerations for the material parameters,
but we show that the principle still stands and a pseudocanalizing
system performs as well as a the canalizing system. Furthermore, this
pseudocanalizing system could be used to improve dark-field hyperlens
designs in terms of the image quality.

\begin{acknowledgments}
This work has received support from Archimedes Foundation (Kristjan
Jaak scholarship) and Villum Fonden (DarkSILD project).
\end{acknowledgments}

\appendix

\section{Derivation of the dispersion relation for HMMs\label{sec:Derivation-of-dispersion}}

Since the system is invariant in $z$-direction we can effectively
consider a two-dimensional case and limit the derivation to TM waves
(for TE waves we would end up with isotropic dispersion). The procedure
in Ref.~\onlinecite{WavePropagationBook} can be followed for general derivation.
We start with

\begin{eqnarray}
\mathbf{E} & = & E_{x}\mathbf{\hat{x}}+E_{y}\hat{\mathbf{y}}\,,\\
\mathbf{H} & = & H_{z}\hat{\mathbf{z}}\,.
\end{eqnarray}
With the help of the constitutive equations we write $\mathbf{D}$
and $\mathbf{B}$ fields as

\begin{eqnarray}
\mathbf{D} & = & \varepsilon_{0}\left(\varepsilon_{x}E_{x}\mathbf{\hat{x}}+\varepsilon_{y}E_{y}\hat{\mathbf{y}}\right)\,,\\
\mathbf{B} & = & \mu_{0}\mu H_{z}\hat{\mathbf{z}}\,.
\end{eqnarray}
After combining the above relations and (source-free) Maxwell's curl
equations

\begin{eqnarray}
\nabla\times\mathbf{E} & = & -i\omega\mathbf{B}\,,\label{eq:maxwell_curl1}\\
\nabla\times\mathbf{H} & = & i\omega\mathbf{D}\,,\label{eq:maxwell_curl2}
\end{eqnarray}
and using $\nabla\cdot\mathbf{D}=0$ we end up with (for $i=x,y$)

\begin{equation}
\frac{1}{\varepsilon_{y}} \frac{\partial^{2}E_{i}}{\partial x^{2}}+\frac{1}{\varepsilon_{x}}\frac{\partial^{2}E_{i}}{\partial y^{2}}+\omega^{2}\varepsilon_{0}\mu_{0}\mu E_{i}=0\,.
\end{equation}

After substituting the plane wave solution $\mathbf{E}=\mathbf{E}_{0}\exp\left(i\mathbf{k}\cdot\mathbf{r}\right),$
we derive the dispersion equation:

\begin{equation}
k_{x}^{2}/\varepsilon_{y}+k_{y}^{2}/\varepsilon_{x}=\mu\omega^{2}/c^{2}\,.\label{eq:derived_dispersion}
\end{equation}
We point out that the magnetic permeability can be anisotropic {[}$\hat{\mu}=\mathrm{diag}\left(\mu_{x},\mu_{y},\mu_{z}\right)${]},
but since one component ($H_{z}$) is nonzero, only $\mu_{z}$ component
would enter the dispersion relation for TM waves.

\section{Phase accumulation for cylindrical waves\label{sec:Cylindrical-theory...}}

We use Maxwell's curl equations {[}Eqs. (\ref{eq:maxwell_curl1})
and (\ref{eq:maxwell_curl2}){]} to obtain the wave equation

\begin{equation}
\nabla\times\left(\hat{\varepsilon}^{-1}\nabla\times\mathbf{H}\right)=k_{0}^{2}\mu\mathbf{H}\,,
\end{equation}
for which we assume TM fields of form $\mathbf{H}\left(r,\theta\right)=F\left(r\right)\exp\left(im\theta\right)\hat{\mathbf{z}}$.
Solving the differential equation yields a general solution for the
fields $\mathbf{H}=H_{z}\mathbf{\hat{z}}$ in the form

\begin{equation}
H_{z}\left(r,\theta\right)=\exp\left(im\theta\right)\left[aH_{\nu}^{(1)}\left(k_{r}r\right)+bH_{\nu}^{(2)}\left(k_{r}r\right)\right]\,,\label{eq:cylindrical_solution}
\end{equation}
where $\nu=m\cdot\sqrt{\varepsilon_{\theta}/\varepsilon_{r}}$ and
$k_{r}=k_{0}\sqrt{\varepsilon_{\theta}}$. The angular momentum mode
number\emph{} $m$ is the number of wavelengths per angle of full
rotation ($2\pi$). By analogy with plane waves, it is useful to introduce
a tangential wavenumber $k_{\theta}$ (i.e. the number of wavelengths
per unit length), which is related to the angular momentum mode number
by $m=k_{\theta}r$. As $m$ is fixed for a given wave component,
we have $k_{\theta}'r_{0}=k_{\theta}r$, from which it follows that

\begin{equation}
k_{\theta}=k_{\theta}'\frac{r_{0}}{r}\,,\label{eq:ktheta_scaling}
\end{equation}
which shows that the wavevectors are compressed during propagation,
corresponding to the magnification of the image.

\begin{figure}
\begin{centering}
\includegraphics{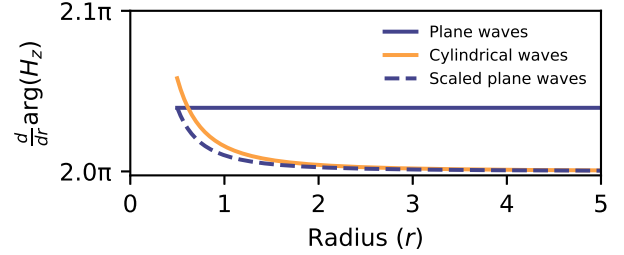}
\par\end{centering}

\caption{\label{fig:phase-accumulation}Acquired phase per distance ($\frac{d}{dr}\mathrm{arg}\,H_{z}$).
For plane waves (solid blue line) the phase is linearly dependent
on propagated distance {[}$\mathrm{arg}\,H_{z}=k_{y}\left(x_{2}-x_{1}\right)${]}.
However, in cylindrical geometry the phase propagation is not linear
--- as seen for Hankel waves (solid yellow line) the phase propagation
depends on the propagation distance. Dashed blue line shows phase
propagation using scaled plane waves {[}Eq.~(\ref{eq:phase_integration}){]},
which offers a good approximation for cylindrical waves.}
\end{figure}

As for a planar system, the aim here is to obtain expressions for
phase accumulation. However, obtaining the analytical expressions
in cylindrical basis is not straightforward. Instead we approach the
problem with modified plane waves. The key difference between planar
and cylindrical geometry is magnification of the image {[}i.e. scaling
of the $k_{\theta},$ given by Eq.~(\ref{eq:ktheta_scaling}){]}.
In planar geometry the phase propagation is expressed with

\begin{equation}
E_{2}=E_{1}\exp\left(ik_{y}\left(x_{2}-x_{1}\right)\right)\,,
\end{equation}
where the acquired phase is just $\mathrm{Re}\left(k_{y}\right)\left(x_{2}-x_{1}\right)$.
To account for magnification arising from cylindrical geometry we
can express the acquired phase by 
\begin{equation}
\mathrm{Re}\intop_{r_{1}}^{r_{2}}k_{r}\left(r\right)\,dr\,.\label{eq:phase_integration}
\end{equation}
In our case $k_r$ is the propagation constant and a function of $r$
(via $k_{\theta}$). The propagation constant $k_{y}$ is given by
Eq.~(\ref{eq:propagation_constant}), however here $k_{x}$ is scaled
as given by Eq.~(\ref{eq:ktheta_scaling}). As shown in Fig.~\ref{fig:phase-accumulation},
this approach of using plane waves with scaled $k_{\theta}$ manages
to capture the phase accumulation effects in cylindrical geometry.
Unlike solution in cylindrical basis, this approach allows for analytical
integration of Eq.~(\ref{eq:phase_integration}).

We anticipate that the material parameters of the second medium must
be scaled to counteract the effect of magnification. We note that
compression of dispersion relation {[}Eq.~(\ref{eq:dispersion}){]}
in $k_{\theta}$ direction is achieved by scaling $\varepsilon_{r}$.
Therefore we alter the second medium as follows: $\varepsilon_{r}^{(2)}=\varepsilon_{r}^{\prime(2)}/\xi$.

Integration of Eq.~(\ref{eq:phase_integration}) can be carried out
analytically, resulting in an analytical expression for the total
acquired phase as a function of the wavevector $k_{\theta}$, geometric
magnification factor $M=1+h/r_{1}$ (where $r_{1}$ is the inner radius
and $h$ is the hyperlens thickness) and scaling parameter $\xi$.
Taking the limit $k_{\theta}\to\infty$ allows us to obtain a simple
expression for material scaling:

\begin{equation}
\xi=\left\{ \ln\left[\left(M+1\right)^{2}/4\right]/\ln\left[\left(M+1\right)^{2}/4M^{2}\right]\right\} ^{2}\,.
\end{equation}

For small magnification factors (i.e. $h\sim r_{1}$ ) the expression
further simplifies to 

\begin{equation}
\xi\approx M\,.\label{eq:approx_scaling}
\end{equation}

It is easy to verify that the resulting expression {[}Eq.~(\ref{eq:approx_scaling}){]}
is a good approximation for our geometry. Furthermore, we numerically
verified that the resulting scaling parameter is the optimal choice
when operating in a proper cylindrical basis using Hankel functions
{[}Eq.~(\ref{eq:cylindrical_solution}){]}.

\end{document}